# A Shift Test for Independence in Generic Time Series

Kenneth D. Harris[1*]

[1]UCL Institute of Neurology, Queen Square, London WC1N 3BG, UK. *Correspondence: kenneth.harris@ucl.ac.uk.

We describe a family of conservative statistical tests for independence of two autocorrelated time series. The series may take values in any sets, and one of them must be stationary. A user-specified function quantifying the association of a segment of the two series is compared to an ensemble obtained by time-shifting the stationary series -N to N steps. If the series are independent, the unshifted value is in the top m shifted values with probability at most m/(N+1). For large N, the probability approaches m/(2N+1). A conservative test rejects independence at significance α if the unshifted value is in the top α(N+1), and has half the power of an approximate test valid in the large N limit. We illustrate this framework with a test for correlation of autocorrelated categorical time series.

Testing the independence of two time series is challenging when the series are themselves autocorrelated (Box, 2008; Granger and Newbold, 1974; Haugh, 1976; Phillips, 1986; Yule, 1926). Most solutions to the problem focus on removing the correlations within a single series, for example by fitting linear autoregressive models and applying standard tests to the residuals. However, this approach is not applicable in many cases. For some data types, such as categorical time series or point processes, decorrelation methods may not be available; furthermore, if the decorrelation methods are inaccurate (such as autoregression for nonlinear time series) they may yield excess false positive errors.

Here we describe a family of independence tests for timeseries taking values in arbitrary sets, one of which is assumed stationary. A user-supplied function quantifies the association of a segment of the two series, and the test compares its value to an ensemble obtained after time-shifting the stationary series. We prove that whatever association measure is used, the test is conservative: it will not falsely reject the null at greater than the nominal rate. An approximate test, with twice the power, is valid in a large N limit.

## The test framework

Consider two time series sampled over $T$ discrete steps: $(X_t)_{t=0}^{T-1}$ and $(Y_t)_{t=0}^{T-1}$, where $X_t$ and $Y_t$ take values in arbitrary sets $\mathbb{X}$ and $\mathbb{Y}$. A user-specified function $V: \mathbb{X}^D \times \mathbb{Y}^D \to \mathbb{R}$ quantifies the association of $X$ and $Y$ on a segment of $D$ consecutive time steps. This function can be arbitrary, but will often be of the loss function of a classifier predicting one series from the other. For real-valued series we might for example use the Pearson or Spearman correlation of $X_t$ with $Y_t$ over a segment of length $D$.

The logic of the test is that it is easier to predict one timeseries from a simultaneous segment of the other, than from a temporally offset segment. We therefore extract a central segment of $X$, $X[N:T-N]$, where $N = (T-D)/2$ and the "Pythonic" notation $X[n:m]$ means $(X_t)_{t=n}^{m-1}$. We quantify the association at shift 0 as $V_0 = V(X[N:T-N], Y[N:T-N])$ and compare this to an ensemble after time-shifting Y: $V_s = V(X[N:T-N], Y[s+N:s+T-N])$, for $s = -N \ldots N$ (**Figure 1**). If $X$ and $Y$ are independent and $Y$ is stationary, then the probability distribution of $V_s$ will not depend on $s$.

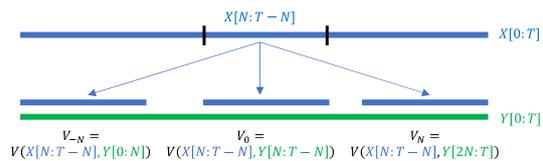

**Figure 1 | Test framework.** The central segment of $X$ is compared to segments along the length of $Y$. If $X$ and $Y$ are independent and $Y$ is stationary, then the distribution of $V_s$ does not depend on $s$.

One might first think to reject the null hypothesis at significance level $\alpha$ if $V_0$ exceeds the $\alpha^{th}$ quantile of the ensemble $\{V_s: s = -N \ldots N\}$. This is valid in the $N \to \infty$ limit, but for any finite $N$ there are examples where it rejects a valid null at twice the nominal significance level (Appendix).

Nevertheless, we can obtain a version of the test that is conservative for any finite $N$. We compute a test statistic $m = \sum_{s=-N}^{N} I(V_s \geq V_0)$, where the indicator function $I$ is 1 if its argument is true, 0 if false; $m$ counts how many shifts produce an association measure as least as big as the unshifted data. If the null is true, then we prove that $\mathbb{P}(m \leq M) \leq M/(N+1)$ (Appendix). To obtain a conservative test at significance level $\alpha$, we thus reject the null if $m \leq \alpha(N+1)$. An approximate version of the test rejects the null if $m \leq \alpha(2N+1)$. This has twice the statistical power but may incorrectly reject the null for small $N$.

### Example

We illustrate the test framework with an example of categorical time series. To simulate independent autocorrelated series (**Figure 2**, left), each independently switches to a random state with probability of 0.1 on each time step. To simulate correlated series (**Figure 2**, right), they also both switch to a common random state with probability of 0.1 per time step. In both cases, series of length 300 were simulated 1000 times. Applying Fisher's exact test to a $2 \times 2$ contingency table for the combinations of $X_t$ and $Y_t$ usually incorrectly rejects independence even if it is true, since the series are autocorrelated (**Figure 2B**).

To apply our framework, we took $N = 19$, allowing the null to be conservatively rejected at p=0.05 if $m = 1$. The relationship between segments of $X$ and $Y$ was quantified by accumulating a contingency table $c_{ij}$ over

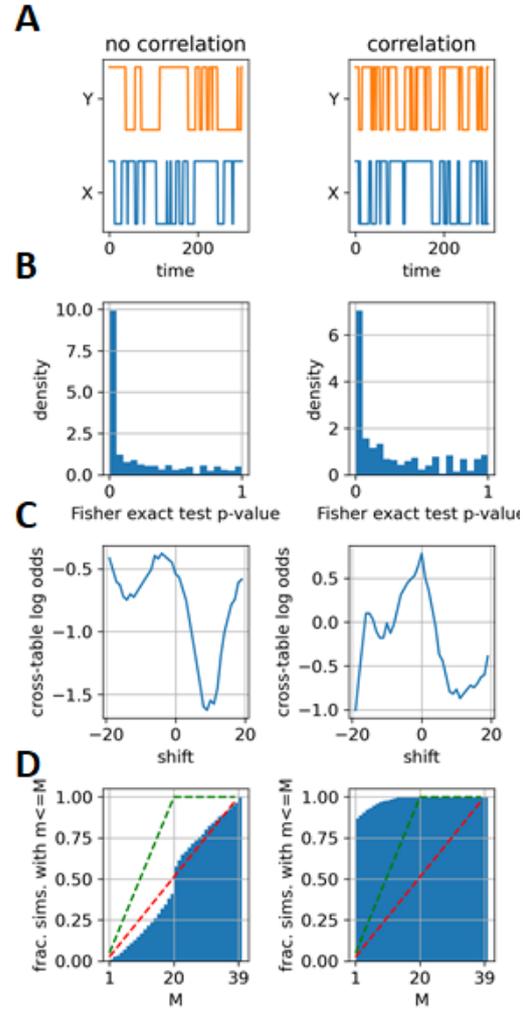

**Figure 2 | Example application to categorical time series. A:** binary timeseries were simulated as independent or correlated Markov chains (left/right). **B:** Histogram of p-values from Fisher's exact test applied to a contingency table of values of $X_t$ and $Y_t$, over 1000 simulations. Due to autocorrelations, the null is usually rejected even when the series are independent (left). **C:** Association between the series was quantified with $V_s$ as the log-odds ratio of a regularized contingency table, after time-shifting $Y$. A peak at shift 0 is seen only for correlated series (right). **D:** cumulative histogram of the test statistic $m = \sum_{s=-N}^{N} I(x_{t+s} \geq x_t)$. Green and red dotted lines show the conservative and approximate bounds for $\mathbb{P}(m \leq M)$.

segments of $X$ and $Y$ of length $D = T - 2N = 262$, and defining $V(X, Y)$ to be

$$V(X, Y) = \log\left(\frac{(\epsilon + c_{00})(\epsilon + c_{11})}{(\epsilon + c_{01})(\epsilon + c_{10})}\right)$$

where $\epsilon = 0.1$ regularizes the case that any of the counts are 0.

Examining the dependence of $V_s$ on $s$ for an example simulation showed smooth dependence for independent series (**Figure 2C**, left), but a sharp peak at $s = 0$ for correlated series (**Figure 2C**, right). We repeated the simulation 1000 times and plotted a cumulative histogram of the test statistic $m$, together with the conservative and approximate bounds for $\mathbb{P}(m \leq M)$ (**Figure 2D**). For uncorrelated series, the conservative bound always held, and the null was rejected because $m \leq 1$ only 10 times out of 1000 simulations, safely below the nominal significance level of 0.05 as well as the significance level of the approximate test, $\frac{1}{2N+1} = 0.026$ (**Figure 2D**, left). When a genuine correlation was present (**Figure 2D**, right) we found that $m \leq 1$ in 869 of 1000 simulations.

Two other features of the null histogram are notable. A sharp jump at $m = 20$ (**Figure 2D**, left) reflected cases where $V_s$ was a monotonic function of $s$. A small jump at $m = 39$ indicated an excess fraction of cases where $V_0 \leq V_s$ for all $s$, due to occasional ties. The conservative bound always held, but the approximate bound was violated near $m = 20$.

### Discussion

We have described a simple family of tests for independence between two time series. We proved that if one of the series is stationary, then the test is conservative, i.e. the probability that it rejects a correct null is never greater than the nominal significance level.

The test relies on a function $V(X, Y)$ that quantifies the association between segments of the two time series. The test will work if $V$ is larger for truly simultaneous segments, than temporally offset segments. Although a poor choice of this function cannot give erroneous rejections of a true null, false nulls will only be rejected if this function is chosen judiciously, based on prior knowledge of the type of dependence expected.

If the two series are suspected to have an instantaneous correlation, then a $V$ comparing simultaneous samples is appropriate. This includes the example above, as well as Pearson or Spearman correlations for real-valued series. With such a $V$, the peak of $V_s$ is expected exactly at $s = 0$, so a small value of N can be used as in our example. If the time series are expected to exhibit less temporally precise correlations, then $V$ might measure the error of predicting $Y_t$ from several neighboring timesteps of $X$; in this case a larger value of $N$ may be required, allowing the null to be rejected if the peak is near but not exactly at $s = 0$. Temporally decorrelating the series may increase statistical power but is not required to avoid false positives. Finally, any data the analyst uses to help select the function $V$ should not be used to assess significance.

The approximate version of the test has twice the power of the conservative test, but this modest gain may not be worth the risk of false conclusions. We suggest that the simplicity of this test framework, together with its conservatism and applicability to generic series, may see it find applications in multiple fields of research.

# Appendix

This appendix contains proofs of the conservative and approximate bounds for $\mathbb{P}(m \leq M)$. We start with the conservative bound $\mathbb{P}(m \leq M) \leq M/(N+1)$, proved as the corollary to Theorem 2.

**Definition.** Let $(x_t)_{t \in \mathbb{Z}}$ be a two-sided infinite sequence of real numbers. We say $t$ is an $(M, N)$-*local maximum* of this sequence if $x_t$ is one of the top $M$ values within $N$ timesteps of $t$:

$$\sum_{s=-N}^{N} I(x_{t+s} \geq x_t) \leq M.$$

Here $I$ is the indicator function: 1 if its argument is true, 0 if false. Clearly, if $t$ is an $(M, N)$-local maximum, it is also a $(M', N)$-local maximum for any $M' > M$.

**Lemma 1.** *Let $(x_t)_{t \in \mathbb{Z}}$ be a two-sided sequence of real numbers. The number of $(M, N)$-local maxima within any consecutive segment of length $N + 1$ is at most $M$.*

**Proof.** For any times $t_i$ and $t_j$ within a length N+1 consecutive segment, $|t_i - t_j| \leq N$. Suppose $\{t_i\}_{i=1}^{M'}$ are a set of $(M, N)$-local maxima within this segment, with $M' > M$. Let $t_j$ be a minimal element of this set, so that $x_{t_i} \geq x_{t_j}$ for all $i = 1 \ldots M'$. Then all the $M'$ values of $t_i$ satisfy $|t_i - t_j| \leq N$ and $x_{t_i} \geq x_{t_k}$ contradicting the assumption that $t_k$ is an $(M, N)$-local maximum. ∎

**Theorem 2.** *Let $(V_t)_{t \in \mathbb{Z}}$ be a real-valued stationary stochastic process. Then the probability that time 0 is an $(M, N)$ local maximum is at most $\frac{M}{N+1}$. This bound cannot be improved: for any $\epsilon > 0$, there exists a process for which the probability exceeds $\frac{M}{N+1} - \epsilon$.*

**Proof.** Define $A_t$ to be a random variable taking the value 1 if $V_t$ is an $(M, N)$-local maximum and 0 if not. $\mathbb{E}[A_0]$ is the probability that 0 is an $(M, N)$-local maximum, and by stationarity, $\mathbb{E}[A_0] = \mathbb{E}[A_t]$ for all $t \in \mathbb{Z}$. Thus,

$$\mathbb{E}[A_0] = \frac{1}{N+1}\sum_{t=0}^{N} \mathbb{E}[A_t] = \frac{1}{N+1}\mathbb{E}\left[\sum_{t=0}^{N} A_t\right]$$

By Lemma 1, $\sum_{t=0}^{N} A_t \leq M$ surely. Thus, $\mathbb{E}[A_0] \leq \frac{M}{N+1}$, providing the required bound.

An example shows the bound cannot be improved. Let $K \in \mathbb{N}$ and for $t = 0 \ldots K(N+1) - 1$ define

$$W_t = \begin{cases} 1 + (t \bmod (N+1)) + M(t \text{ div } (N+1)) & t \bmod (N+1) < M \\ 0 & t \bmod (N+1) \geq M \end{cases}$$

where *mod* and *div* mean integer modulus and division. For example, if $M = 3, N = 4, K = 3$ then $(W_t) = (1,2,3,0,0,4,5,6,0,0,7,8,9,0,0)$. Every $t$ with $W_t > 0$ is an $(M, N)$ local maximum of this sequence.

Now let $(V_t)_{t\in\mathbb{Z}}$ be the stationary stochastic process obtained by concatenating an infinite number of copies of $(W_t)$, shifted by a random integer uniformly distributed in $[0, K(N+1))$. Every $t$ with $V_t \geq M$ is an $(M, N)$-local maximum and there are $MK - M + 1$ of in every $K(N+1)$ timesteps, so by stationarity the probability that 0 is an $(M, N)$-local maximum is $\frac{MK-M+1}{K(N+1)}$. Choosing $K > \frac{M-1}{(N+1)\epsilon}$ we obtain that this probability $> \frac{M}{N+1} - \epsilon$. ∎

**Corollary.** Let $(X_t)_{t=0}^{T-1}$ be a stochastic sequence taking values in a set $\mathbb{X}$, and $(Y_t)_{t=0}^{T-1}$ be a segment of a stationary stochastic sequence $(Y_t)_{t\in\mathbb{Z}}$ taking values in a set $\mathbb{Y}$, with $(Y_t)$ independent of $(X_t)$. Let $N$ be an integer $< T/2$ and let $V: \mathbb{X}^D \times \mathbb{Y}^D \to \mathbb{R}$ be a measurable function, where $D = T - 2N$. For $s = -N \ldots N$ let $V_s = V(X[N:T-N], Y[s+N:s+T-N])$, where $X[a:b]$ means the segment $(X_t)_{t=a}^{b-1}$. Let $m = \sum_{s=-N}^{N} I(V_s \geq V_0)$. Then $\mathbb{P}(m \leq M) \leq M/(N+1)$, for all $1 \leq M \leq N+1$.

**Proof.** Because $(Y_t)_{t\in\mathbb{Z}}$ is stationary and independent of $(X_t)$, $V_s$ is stationary also. So, by theorem 2, the probability that 0 is a $(M, N)$-local maximum of $(V_s)$ is at most $M/(N+1)$. If $m \leq M$ then 0 is a $(M, N)$-local maximum, therefore $\mathbb{P}(m \leq M) \leq M/(N+1)$. ∎

This establishes the conservative test.

The approximate test rejects the null at significance level $\alpha$ if $\sum_{s=-N}^{N} I(V_s \geq V_0) \leq \alpha(2N+1)$. To demonstrate the asymptotic validity of the approximate test we fix $D$ (the length of the segments of X and Y being compared) and let $N \to \infty$. We will show that for $N$ large enough, the false rejection rate exceeds nominal rate by an arbitrarily small amount. We have already established that $(V_s)$ is stationary, so it suffices to establish the following theorem:

**Theorem 3.** Let $(V_s)_{s\in\mathbb{Z}}$ be a stationary sequence and let $0 < \alpha < 1$. Then

$$\limsup_{N\to\infty} \mathbb{P}\left(\sum_{s=-N}^{N} I(V_s \geq V_0) \leq \alpha(2N+1)\right) \leq \alpha$$

**Proof.** We first consider the case that $(V_s)$ is ergodic. Let $\epsilon > 0$ and let $q$ be a $(1-\alpha-\epsilon)^{th}$ quantile of $\mathbb{P}(V_0)$, meaning that $\mathbb{P}(V_0 \geq q) \geq \alpha + \epsilon$ and $\mathbb{P}(V_0 > q) \leq \alpha + \epsilon$; the same holds for all $V_s$ by stationarity. Write $p_N = \mathbb{P}(\sum_{s=-N}^{N} I(V_s \geq V_0) \leq \alpha(2N+1))$. Because the probabilities of disjoint events add:

$$p_N = \mathbb{P}\left(V_0 \leq q \cap \sum_{s=-N}^{N} I(V_s \geq V_0) \leq \alpha(2N+1)\right) + \mathbb{P}\left(V_0 > q \cap \sum_{s=-N}^{N} I(V_s \geq V_0) \leq \alpha(2N+1)\right)$$

Because $\{V_0 \leq q \cap \sum_{s=-N}^{N} I(V_s \geq V_0) \leq \alpha(2N+1)\} \subseteq \{\sum_{s=-N}^{N} I(V_s \geq q) \leq \alpha(2N+1)\}$ and $\{V_0 < q \cap \sum_{s=-N}^{N} I(V_s \geq V_0) \leq \alpha(2N+1)\} \subseteq \{V_0 < q\}$:

$$p_N \leq \mathbb{P}\left(\sum_{s=-N}^{N} I(V_s \geq q) \leq \alpha(2N+1)\right) + \mathbb{P}(V_0 < q)$$

$$\leq \mathbb{P}\left(\sum_{s=-N}^{N} I(V_s \geq q) \leq \alpha(2N+1)\right) + \alpha + \epsilon.$$

By ergodicity, $\frac{1}{2N+1}\sum_{s=-N}^{N} I(V_s \geq q_{\alpha+\epsilon})$ converges in probability to $\mathbb{P}(V_0 \geq q_{\alpha+\epsilon}) \geq \alpha + \epsilon$, and so $\limsup_{N\to\infty} \mathbb{P}(\sum_{s=-N}^{N} I(V_s \geq q_{\alpha+\epsilon}) \leq \alpha(2N+1)) = 0$. Thus,

$$\limsup_{N\to\infty} p_N \leq \alpha + \epsilon.$$

Because $\epsilon$ was arbitrary, this proves the theorem for $(V_s)$ ergodic.

Now consider $(V_s)$ stationary but not necessarily ergodic. Let $Q$ be a $(1-\alpha-\epsilon)^{th}$ conditional quantile with respect to the shift-invariant sigma algebra $\mathcal{I}$, so $Q$ is a $\mathcal{I}$-measurable random variable with $\mathbb{P}(V_s \geq Q \mid \mathcal{I}) \geq \alpha + \epsilon$ and $\mathbb{P}(V_s > Q \mid \mathcal{I}) \leq \alpha + \epsilon$ almost surely. There always exists at least one such $Q$ (Dembińska, 2014; Tomkins, 1975). Therefore, $\mathbb{P}(V_s \geq Q) \geq \alpha + \epsilon$ and $\mathbb{P}(V_s > Q) \leq \alpha + \epsilon$, and the same argument as above yields

$$p_N \leq \mathbb{P}\left(\sum_{s=-N}^{N} I(V_s \geq Q) \leq \alpha(2N+1)\right) + \alpha + \epsilon$$

By the ergodic theorem (e.g. Billingsley, 1995, theorem 24.1; Grimmett and Stirzaker, 2001, theorem 9.5.16), $\frac{1}{2N+1}\sum_{s=-N}^{N} I(V_s \geq Q)$ converges in probability to $\mathbb{E}(I(V_s \geq Q) \mid \mathcal{I}) = \mathbb{P}(V_s \geq Q \mid \mathcal{I})$, which $\geq \alpha + \epsilon$ almost surely. Thus again $\limsup_{N\to\infty} p_N \leq \alpha + \epsilon$ for arbitrary $\epsilon$, completing the proof. ∎

**Acknowledgements:**
This work was supported by the Wellcome Trust (205093), European Research Council (694401), and Simons Foundation.